# Development of travelling heater method for growth of detector grade CdZnTe single crystals

P. Vijayakumar[a], Edward Prabu Amaladass[a,b], K. Ganesan[a,b], R. M. Sarguna[a], Varsha Roy[a], S. Ganesamoorthy[a,b,*]

[a]*Materials Science Group, Indira Gandhi Centre for Atomic Research, Kalpakkam, 603 102, India*
[b]*Homi Bhabha National Institute, Training School Complex, Anushakti Nagar, Mumbai, India*
[*]Email: sgm@igcar.gov.in

**Abstract**

We report on the indigenous design and development of laboratory scale travelling heater method (THM) system to grow detector grade $Cd_{0.9}Zn_{0.1}Te$ (CdZnTe) single crystals. THM system mainly consists of two-zone furnace with a tuneable temperature gradient (30 - 80 $^0$C/cm), high precision translation (1 - 25 mm/day) and rotation (1 - 50 rpm) assemblies to meet the stringent conditions that are essential to grow detector grade CdZnTe single crystals. Further, a load cell in the THM system enables continuous monitoring of the growth. Systematic growth experiments were performed to optimize the various growth parameters in order to achieve large grain single crystals. Herein, the effect of temperature gradient and growth rate on the increase in grain size is discussed in detail. Each successful growth experiment yields a minimum of four detector grade elements of dimensions 10 x 10 x 5 mm$^3$ from a starting charge of 100 g of CdZnTe. The crystalline nature and quality of the detector elements were evaluated using Laue, NIR transmission spectroscopy and I-V characteristics. Crystals with resistivity greater than ~ $10^9$ - $10^{10}$ Ω-cm were identified for testing gamma ray detection. The photo peak of $^{137}$Cs was resolved with an energy resolution of 4.2 % at 662 keV and its measured electron mobility lifetime product is found to be ~ 3.3 x $10^{-3}$ cm$^2$/V. The demonstration of the gamma ray detection with a relatively high µτ product is the testimony to the successful growth of detector grade CdZnTe single crystals by an indigenously developed THM system.

Keywords: Travelling heater method; single crystal growth; CdZnTe; Gamma ray detector; I-V characteristics

# 1. Introduction

Cadmium zinc telluride (CdZnTe) is one of the most promising room-temperature semiconductor gamma radiation detector material due to its wide energy band gap and high atomic number [1–4]. The research activities carried out on detector technology over the past few decades have showcased that CdZnTe is an excellent substitute for scintillation detectors in terms of high sensitivity, better energy resolution and high spatial resolution at room temperature [4–7]. In the recent past, CdZnTe detector with an energy resolution of less than ≈ 0.5 % at 662 keV of $^{137}$Cs had already been demonstrated [7] and commercial detectors are also available in the market. Even though its resolution is inferior to the resolution exhibited by HPGe detectors (0.2 % at 662 keV), the ease of operation with no cooling requirements renders the superiority of CdZnTe detectors for gamma-ray spectrometric applications. Moreover, the pixelated CdZnTe detectors are also becoming indispensable for room temperature detection of hard X-rays in medical and industrial imaging applications [2,7,8]. Despite having several advantages, its high production cost and the complications involved in the crystal growth impose huge challenges to grow large volume detector grade CdZnTe crystals at reasonable cost. Presently, only a few crystal growth industries are involved in the production and commercialization of large size CdZnTe single crystals for gamma radiation detector and spectroscopic imaging applications. But, the availability of detector grade CdZnTe crystals/wafers with homogeneous and low defect density at a reasonable cost is still limited. This leads to an exploration of crystal growth methods at laboratory scale in pursuit to obtain detector grade CdZnTe crystals. However, the successful demonstration of growth of detector grade CdZnTe single crystals by academia is still limited in the literature.

The growth of CdZnTe crystals was initially started with different varieties of Bridgman (BM) techniques , such as vertical BM, horizontal BM and high pressure BM methods [9,10,19,11–18]. Further, the vertical gradient freeze (VGF), travelling heater method (THM) and physical vapour transport were also used for the growth of CdZnTe single crystals. In Bridgman and VGF techniques, the crystal growth is conducted at the melting point of CdZnTe of 1115 ºC which results in continuous loss of Cd, Zn segregation, and compositional inhomogeneity along the growth axis. Also, the melt becomes rich in Te during growth. Further, the Te inclusions with sizes ranging from 5 – 60 μm occur depending upon the growth conditions. Note that the Te precipitates/inclusions, especially those larger than 10 μm, causes electron trapping leading to deterioration in the signal quality and severe degradation of detector performance [20].

The THM is the most successful crystal growth method that has been shown to produce detector-grade CdTe and CdZnTe crystals consistently by different industries including ACRORAD, Co., Ltd, Japan [3,4,13,16–19,21]. The main advantage of THM is the possibility of growing crystals at lower growth temperature due to the addition of excess Te which is used as a solvent. Since the growth temperature is low and only a partial solute is in the molten state, the defect density and Te precipitates can be minimized significantly and also uniform Zn concentration along the radial and axial growth directions can be obtained. However, the significant advantages of THM are hampered by slow growth rate of ~ 1 – 5 mm/day and the formation of multi-grains which limits the possibility of obtaining useable volume of detector grade single crystal from the entire volume of the grown ingot. Hence, one needs to have highly sophisticated crystal growth facility with fine tunability on various process parameters to control the crystal growth kinetics which enables high structural quality and low defect density CdZnTe single crystals. Further, the optimization of THM process parameters leading to single large grain is an extremely challenging task.

In literature, several reports are available on the growth of CdZnTe single crystals by THM. However, it is understood that the growth of high-quality CdZnTe single crystals involves several technical challenges due to the inherent materials properties such as very low thermal conductivity (0.01 W/mK), high segregation co-efficient of Zn ($K_{eff}$ ~1.35) and propensity to defects due to high vapour pressure of Cd and growth instability at the solid-liquid interface. The THM system can overcome some of these challenges with its unique design. The THM system was initially developed for the growth of HgCdTe for infrared imaging applications and later, THM is extensively used for growth of CdTe based II-VI compounds [15–19,22–24].

In this report, we highlight the indigenous design and development of laboratory scale THM system for the growth of small volume CdZnTe crystal in a small diameter quartz ampoule but capable of yielding a large usable volume of detector elements from a single growth run. To achieve this goal, a large number of growth experiments were performed systematically. Herewith, we discuss the effect of temperature gradient and growth rate on the growth of large grain single crystals with a diameter of 20 mm. After the successful growth, the CdZnTe crystals were characterized and the suitable CdZnTe detector elements were tested for the detection of gamma rays from $^{137}$Cs source and the results are discussed here.

## 2. Experimental

### 2.1 Design of THM crystal growth setup

The major components in the THM system are the high precision translation stage, ampoule rotation assembly, growth furnace that encompasses constant temperature zone, high temperature gradient zone and a post growth annealing zone. These components were indigenously custom-designed by us and were fabricated by M/s. Raana Semiconductors Pvt. Ltd., Hosur, India. The block diagram of the developed THM crystal growth system is shown in Fig. 1a and its major components in the growth system are listed in the figure caption with serial numbers. A nearly vibration free translation and rotation of crucible is a crucial part of single crystal growth. The developed THM system is custom designed with precise control options for achieving a translation rate ranging from 0.5 - 25 mm/day with negligible vibrations and capable of transporting a maximum of 25 kg of heater assembly. The required translation rate is achieved using stepper motor with harmonic gear having torque of 5 Nm and step angle resolution of 0.0036° per pulse. In addition, the translation stage can also be moved at higher speeds in the range of 0.1 – 150 mm/min for fast up/down movement of heater assembly using a remote control. Rotation of growth ampoule is managed through a harmonic geared stepper motor of 2.4 Nm torque with step angle resolution of 0.0072° to achieve rotation rate in the range of 0.1 – 50 rpm in both clockwise and counter clockwise directions. Other essential features like electromagnetic braking, digital linear scale (0.01 mm accuracy) for movement monitoring, limit adjustable switches for setting the translation limits, and height adjustable furnace base plate are provided in this system.

The growth unit has a centre chuck for holding the quartz ampoule so that the wobbling of the quartz ampoule can be avoided. Further, the ampoule may crack during the crystal growth process due to either improper growth conditions, or poor vacuum sealing. Furthermore, cracks in quartz tube and the exothermic reaction in the precursor chemicals may also results in a failure run. To overcome these issues, the crystal growth system is provided with a high precision (0.01 mg) load cell assembly through which the weight of the growth ampoule is continuously monitored. The growth quartz ampoule is directly loaded onto the load cell. If any catastrophic failure occurs, it can be immediately recognized by the loss of weight in the load cell. Then, the growth process can be shutdown automatically to minimize chemical toxicity. Due to the toxic nature of the Cd and Te elements, a cabinet with concealed exhaust outlet is provided for the THM assembly. This also helps to minimize the temperature fluctuation during the growth process. The cabin is assembled using aluminum plates, glass windows, and air-tight silicone beading. The controllers of THM system are placed outside of the cabin. A chartless recorder 6100A monitors and records the working temperature using an R-type / K-type thermocouple.

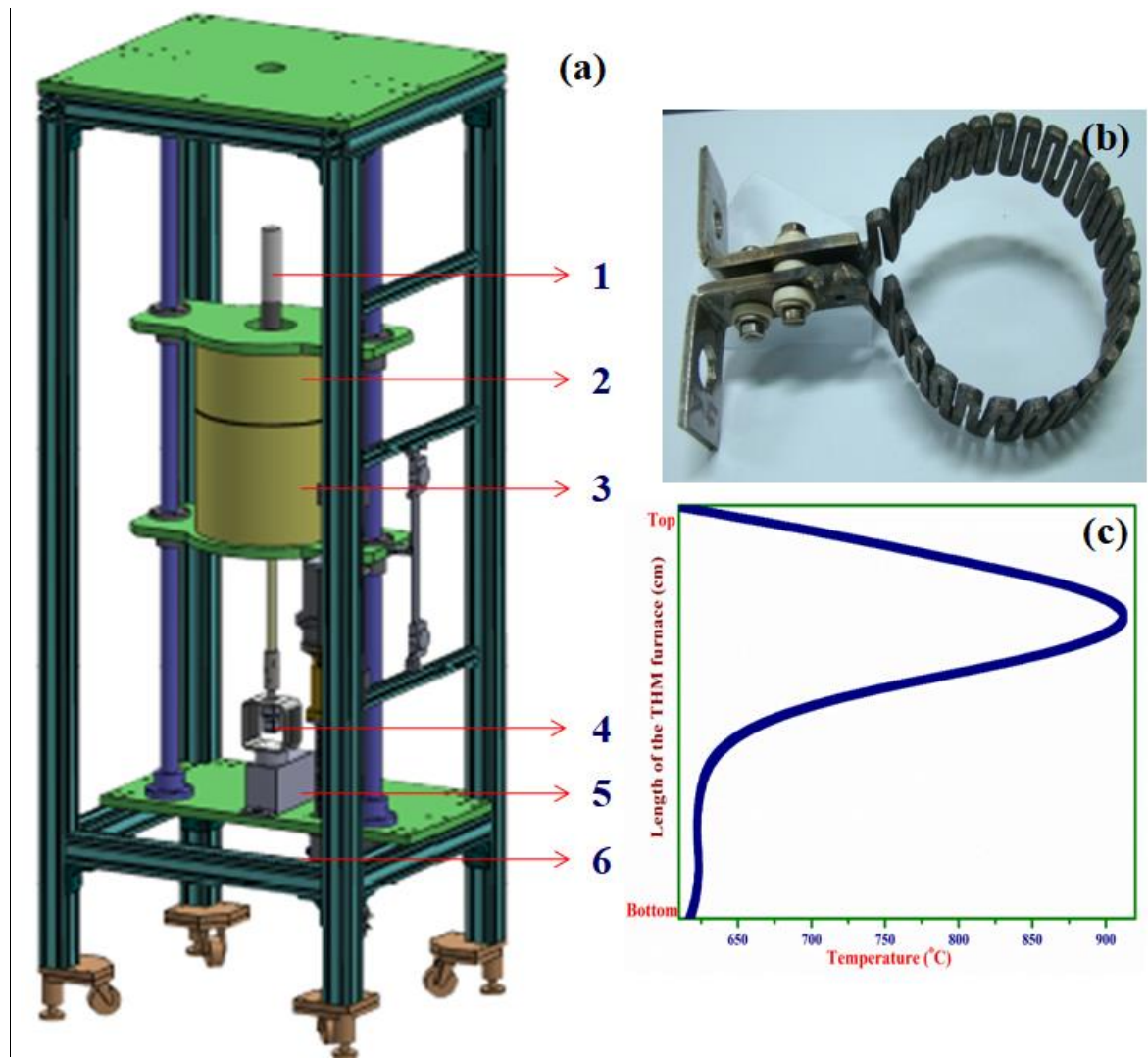

*Fig.1. (a) Schematic of the developed THM crystal growth system. The serial numbers in the figure represents, 1- quartz ampoule, 2- growth heater, 3- annealing heater, 4- ampoule rotation motor, 5- load cell, 6-translation motor, respectively, (b) CNC machined strip heating coil, and (c) typical temperature profile utilized for growth of CdZnTe crystal by THM process*

The basic requirements of the furnaces are an isothermal region of ~ 850 – 950 °C with ~ 10 mm zone length, high temp gradient of 10 - 80 °C/cm in the growth regime and an isothermal temperature region of ~ 600 °C for annealing the grown crystals. The heaters were designed and developed indigenously and a representative resistive strip heater is shown in Fig. 1(b), having the height of 25 mm that was cut and made with computer numerical control (CNC) machine. Then, the CNC cut heater strip was fired by a step down transformer to make it into a circular coil. A typical temperature profile required for CdZnTe crystal growth is shown in Fig.1(c). The height, diameter and shape of the heater coil were optimized to get the isothermal zone. The low thermal conductivity of CdZnTe crystal limits the transport of latent heat during the crystallization and therefore reduces the growth rate. This issue can be overcome to some extent by adopting a large temperature gradient (30 - 50 °C/cm) for crystallization. However, the optimization of appropriate temperature gradient is a challenging task since the large temperature gradient can significantly alter the growth interface due to strong convection in the molten region. The rotation of the growth ampoule is thus introduced to reduce the surface boundary layers near the solid-liquid interface. Depending on the rotation rate, the growth interface shape can be tailored from concave to flat to convex [16–19]. The required temperature gradient was achieved by carefully adjusting the spacer between the growth and the annealing heaters. The heater assembly was placed inside a stainless steel enclosure with appropriate filling materials. The deployed post growth annealing zone has a constant temperature zone length of 50 - 80 mm. Here, a large number of heater assemblies were designed with appropriate liners, diameter, wall thickness, height, spacing between growth and annealing heaters for optimizing the required temperature profile and they were tested before each growth experiments. Furnace control cabinet unit consists of a Eurotherm temperature controller with ramping rate of as small as 0.01 °C/h, thyristors (60 A), step down transformers (4 kVA), safety fuses, current and voltage meters. Each resistive heated zone is independently controlled through the control circuits to achieve the desired temperature gradient.

### 2.2. Synthesis of CdZnTe polycrystalline material

Semiconductor grade quartz ampoules with inner and outer diameter of 19 and 22 mm respectively, were used for the synthesis. Initially, the quartz ampoules were cleaned by chemical methods and stored in a hot oven (~ 90 °C). Subsequently, the quartz ampoules were coated with a thin graphite layer and annealed in vacuum at 1000 °C. Then, the stoichiometric ratio of high pure (7N) elements of Cd, Zn, and Te to yield $Cd_{0.9}Zn_{0.1}Te$ were loaded into the graphite coated quartz ampoule along with excess Te to reduce the synthesis temperature and to maintain stoichiometric composition in the solute. The elements were weighed and loaded into the quartz ampoule using a glove bag pressurized with argon gas.

We followed a similar synthesis process reported in the literature [25,26]. In addition, an appropriate amount of *In* dopant in ppm level was also added to the precursor materials to compensate for Cd vacancy and to achieve high resistivity in the CdZnTe crystals [27,28].

### 2.3. Crystal Growth

The in-house developed THM system was used to grow CdZnTe single crystals. A conically tapered semiconductor grade quartz ampoules were taken for the growth. At first, the requisite Te solvent and the synthesized CdZnTe polycrystalline rod were loaded into a quartz ampoule of 20 mm inner diameter and then, it was evacuated and sealed under the vacuum of $1 \times 10^{-6}$ mbar. The loaded quartz ampoule was placed at the maximum temperature region of the THM crystal growth system. A slow heating of 10 ºC/h is adopted till 600 ºC to melt Te solvent. Then, the furnace is heated up to 850 - 950 ºC depending upon the excess Te solvent taken for the growth. Subsequently the molten zone, the excess Te solvent and the dissolved CdZnTe from the synthesized rod, were maintained at constant temperature for a few hours to homogenize the solution. Then, the heater assembly was translated upward at a very slow movement rate of 1 - 20 mm/day until the crystal growth run was completed. Here, the growth was initiated with spontaneous nucleation occurring at a small volume in the conically tapered quartz ampoule by super cooling process. After completing the growth, the ampoule was quickly moved into the annealing zone and annealed at a temperature of ~ 600 °C for about 48 hours. The furnace was then cooled down to room temperature at 5 – 10 ºC/h. In some growth runs quartz ampoule got cracked during the crystal growth process due to improper growth conditions. In such cases, the THM system will immediately undergo an automatic shutdown process based on the weight loss monitored by load cell assembly through a software control unit. Since each crystal growth runs takes a few weeks for completion, the load cell inputs to the software control unit help to know about the loss of toxic Cd and Te and to prevent accidents.

A crucial factor in the growth of detector grade CdZnTe is the quantity of the Te solvent for the chosen quartz tube diameter that needs to be optimized. Other important factors such as growth temperature, temperature gradient, growth rate, and rotation rate also need to be optimized for the given diameter of the quartz ampoule. Based on the phase diagram of $Cd_{0.9}Zn_{0.1}Te$ and existing literature [13,16–19,29–31], a set of initial parameters such as Te solvent concentration, growth temperature, growth rate and temperature gradient were chosen for single crystal growth using THM. Subsequently, several trial growth experiments were performed by systematically varying each parameter or a combination of parameters for optimizing the growth of CdZnTe single crystals. After each growth run, the crystal was cut perpendicular to the growth axis and its microstructure was analysed with optical microscopy. Since the primary goal was to achieve the large grain single crystal, the cross-sectional optical microscopic images helped to optimize the growth parameters. Thus, the effects of the process parameters on the grain size of CdZnTe crystals were assessed immediately with optical microscopy. Fig. 2 shows the photograph of successfully grown CdZnTe crystal having diameter and length of 20 and 70 mm, respectively. The as-grown CdZnTe crystals are usually covered with a thin Te layer which was removed subsequently. Further, the grown crystal's external surface is found to be smooth with only a few pits on them. Then, these grown crystals were diced to observe the grains size. In the early growth runs, the grown crystals comprise mostly polycrystalline structures with small grain sizes. After rigorous optimization of the process parameters, an increase in grain size was observed. The typical process parameters used for optimization in the growth of 100 g of initial charge in a quartz ampoule of 20 mm inner diameter as follows: i) Te concentration of 20-30 g, ii) temperature gradient of 20 - 50 °C/cm, iii) growth rate of 2 - 5 mm/day and iv) rotation rate of 5 - 10 rpm. These growth parameters are also mutually dependent on each other, and they have to be optimized together appropriately.

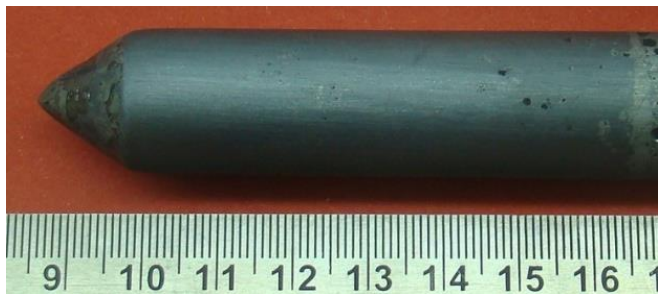

*Fig. 2. Photograph of a typical $Cd_{0.9}Zn_{0.1}Te$ crystal grown by THM system*

### 2.4 Effect of temperature gradient and growth rate on the CdZnTe crystals

Figures 3a - 3d show the photographs of the cross-sectional view of CdZnTe crystals grown at the temperature gradient of 20, 30, 40 and 50 °C/cm respectively, while the growth rate was fixed at 5 mm/day. These wafers were cut at a height of ~ 60 % from the bottom of the crystal for comparison of grain size. At a low-temperature gradient of 20 $^0$C/cm, the crystal consists of polycrystalline structure with very small grain sizes as shown in Fig. 3a. As the temperature gradient increased to 30 °C/cm, the number of grains decreases and the grain size increases significantly as evidenced in Fig. 3b. Further increase in temperature

gradient (40 °C /cm), the grain size becomes large with only four observable grains in the crystal (Fig. 3c). At the temperature gradient of 50 °C /cm, the large grain CdZnTe crystal is grown with only two number of grains in it (Fig. 3d). Thus, the increase in grain size is obtained with increasing temperature gradient from 20 to 50 °C /cm for a constant growth rate of 5 mm/day. Further, growth optimization was continued by varying the growth rate while fixing the temperature gradient and other process parameters.

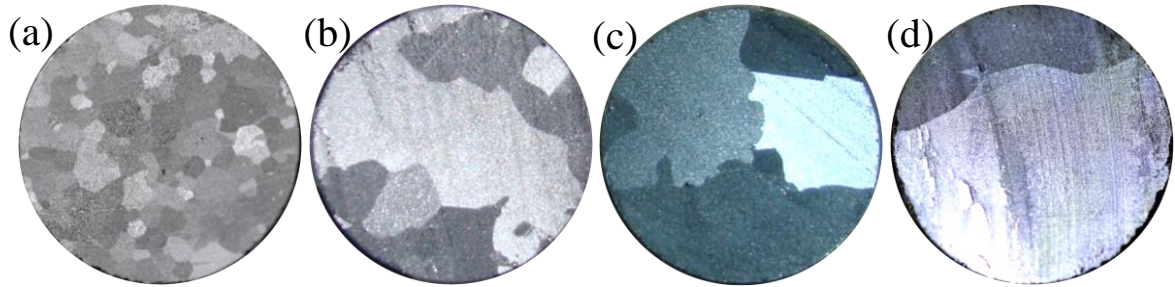

*Fig. 3. Photographs of the cross-sectional view of CdZnTe wafers that are grown at the temperature gradient of (a) 20, (b) 30, (c) 40 and (d) 50 °C /cm while all other parameters were kept constant. The diameter of all these wafers is 20 mm.*

Figures 4a-4d show the cross-section optical micrograph of CdZnTe wafers which were grown at growth rate of 2, 4, 6, and 8 mm/day respectively, while the temperature gradient was at 45 °C /cm and other parameters were kept constant. Here, each wafer was cut perpendicular to the growth axis at ~ 60 % of the height from the bottom of each CdZnTe ingot for microstructure comparison. For the temperature gradient of 45 °C /cm, the crystal is found to have multigrain structure at the growth rate of 2 and 8 mm/day as evidenced by optical micrographs shown in Figs. 4a and 4d, respectively. However, the crystal is found to be single crystal when the growth rate was kept at 4 mm/day, as shown in Fig. 4b. Also, the grown crystal is found to be nearly single crystal with an additional small grain when the growth rate was at 6 mm/day. Thus, the temperature gradient of 45 °C /mm and growth rate of 4 mm/day enable CdZnTe crystal to have a large single grain upto 95 % of the ingot.

Here, we provide a brief account on the growth mechanism of CdZnTe crystals in the THM. The temperature gradient at the growth interface plays a very crucial role in the growth of CdZnTe single crystals since it can alter the growth kinetics and thermodynamic equilibrium in the THM. The maximum permissible growth rate depends on the mass transport through the Te+CdZnTe solution zone which allows the growing interface to advance. In THM system, the maximum growth rate ($v_{cs}$) under equilibrium condition can be written as [24],

$$v_{cs} = \frac{Dc_L \Delta H_s G_L}{(c_{sg}-c_{lg})RT_L^2} \quad \ldots\ldots\ldots (1)$$

Here, $G_L$ is the temperature gradient, $\Delta H_s$ – latent heat of the crystallization, $c_{sg}$ and $c_{lg}$ - concentration of CdZnTe in the solid and liquid phases near the growth interface, $c_L$ – the average concentration of CdZnTe in the liquid phase, D – diffusivity of CdZnTe in the solvent Te, and $T_L$ – the absolute temperature at the liquid-solid interface. Here in THM, the molten zone which consists of CZT solute and Te solvent moves across the pre-synthesized CZT feed. Among the two liquid-solid interfaces, the crystallization occurs at the receding side and dissolution of feed occurs at the other side. The ratio of Cd to Te dictates the molten zone temperature. At the solid-liquid growth interface, excess tellurium is ejected from the crystallized CdZnTe surface and it diffuses back into the solution zone. This creates a layer with excess Te near growth interface while there is depletion of Te layer at the other interface caused by dissolution of CdZnTe feed. Diffusion of excess Te transports upward direction across the molten zone whilst the CdZnTe solute transports towards growth interface [32]. Thus, the growth rate is limited by the transport of solute and solvent which depends on the temperature gradient, growth rate and diffusivity of the species in the molten zone. Moreover, the thermal conductivity of solid CdZnTe is very low of ~ 0.035 W/cm.K and hence, the removal of latent heat associated with crystallization is poor. Hence, the crystal tends to become multigrain structure at the higher and lower growth rates.

In a typical experimental conditions, the maximum growth rate has to be less than 10 mm/day for the temperature gradient of 70 °C /cm [24,33]. If the growth rate is higher than that of 10 mm/h, the macroscopically smooth interface converts into a cellular substructure near the growth interface [33]. Such cellular substructure forms due to the fact that the actual temperature at some point in front of the interface is lower than the equilibrium temperature of the solution for the actual concentration of the species. This also leads to a concentration gradient near the interface, eventually changing the local freezing point of solute within the liquid zone. Thus, the concentration gradient increases at higher growth rates for a given temperature gradient, leading to supercooled liquid in front of the interface. Moreover, this growth instability near the interface results in the formation of Te precipitates. To minimize such cellular substructure and Te precipitates, the temperature gradient was varied from 20 to 50 °C /cm while the other process parameters were kept constant. According to the classical theory of constitutional supercooling, a large axial temperature gradient is believed to suppress constitutional supercooling and allow increased growth rate. Moreover, a large longitudinal temperature gradient is easier to be realized and minimizes the effects of temperature fluctuation near the growth interface [34]. Thus, the increase in temperature gradient helps in the growth larger grain size which eventually minimizes the number of grains. Here, the growth rate of 4 mm/day is found to be optimal for growing CdZnTe single crystals at the temperature gradient of 45 °C /cm at the growth interface.

From a starting charge of 100 g of CdZnTe, each successful crystal growth run yields atleast 4 detector elements with dimensions of 10 x 10 x 5 mm$^3$. After establishing the growth of large grain CdZnTe crystals, the wafers were cut and surface processed before being taken for analysis. Fig. 5a depicts the four surface processed wafers taken from one CdZnTe crystal. Then, these surface processed wafers underwent various quality assessments to ascertain their structural, optical, and electrical properties which are discussed below. Fig. 5b shows the quasi-hemispherical detector prepared by gold electrode on the five sides of the CdZnTe wafer by the electroless process.

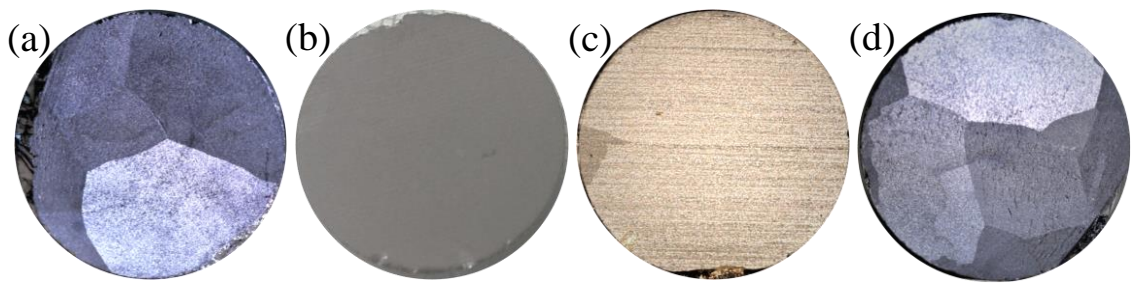

*Fig. 4. Photographs of cross-sectional view of CdZnTe wafers that are grown at the growth rate of (a) 2, (b) 4, (c) 6 and (d) 8 mm/day while all other parameters were kept constant. The diameter of all these wafers is 20 mm.*

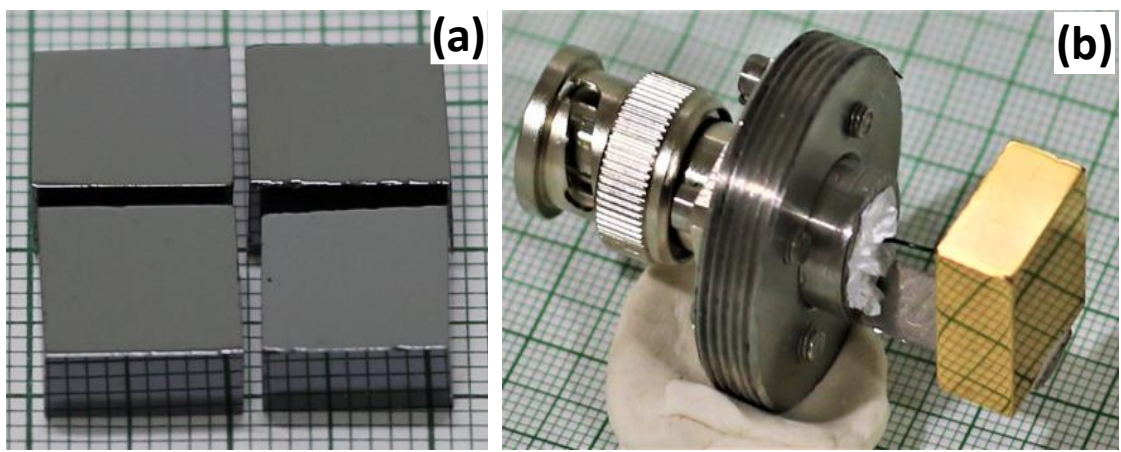

*Fig. 5. (a) Surface polished CdZnTe wafers and (b) A quasi-hemispherical detector prepared by gold electrode on the five sides of the CdZnTe wafer by the electroless process.*

## 3. Results and Discussion

### 3.1 X-ray diffraction

Figure 6 shows the powder X-ray diffraction pattern measured on the CdZnTe wafer displaying a single peak corresponding to the growth orientation of <111>. Further, backscattered Laue diffraction studies also affirmed the single crystalline nature of the THM grown CdZnTe wafers. The inset of Fig. 6 shows the recorded Laue diffraction pattern of a typical CdZnTe wafer. A symmetric and periodic arrangement of diffraction spots confirms the single crystalline nature of grown CdZnTe crystal. The three fold symmetry in the Laue pattern indicates the orientation of the CdZnTe wafer to be (111).

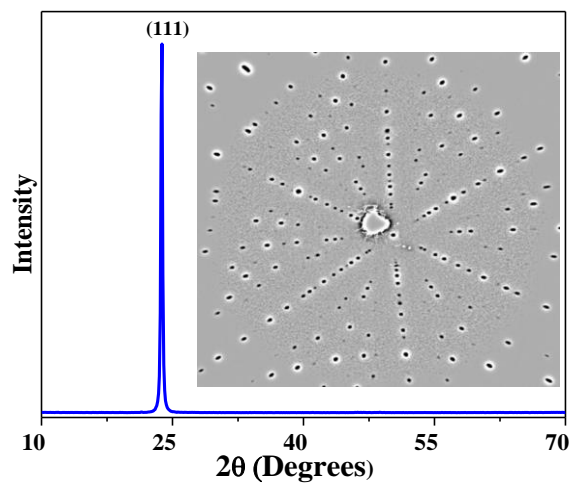

*Fig. 6 X-ray diffraction pattern of CdZnTe wafer indicating the <111> orientation of the grown single crystal and the inset shows the Laue diffraction pattern*

### 3.2. Infrared optical transmission spectroscopy

Figure 7 displays the NIR transmission spectra of the 2 mm thick double-side polished CdZnTe wafers that are taken at bottom, middle and top portions of the crystal. Bruker vertex 80V spectrometer with resolution of 4 cm$^{-1}$ and aperture size of 3 mm was used for the measurement. The inset in Fig. 7 shows the Tauc plot for these wafers. As can be seen from Fig. 7, the estimated optical band gap is very uniform of ~ 1.52 eV for the CdZnTe wafers from different part of the crystal. Further, the Zn composition in CdZnTe crystal can also be extracted from NIR absorption spectra. It is well known that the variation of the band gap of a ternary alloy can be written by the quadratic equation as given below, [35]

$$E_g^{Cd_{1-x}Zn_xTe}(x) = (1-x)\, E_g^{CdTe} + x\, E_g^{ZnTe} - bx(1-x) \qquad (2)$$

Here, $E_{CdTe}$ and $E_{ZnTe}$ are optical band gap of CdTe and ZnTe crystal respectively, b is the bowing factor of 0.46 and x is the Zn composition. Based on the equation 2, the estimated Zn concentration is ~ 0.099 at % which is uniform throughout the crystal from bottom to top. Also, the observed maximum transmittance of ~ 60 % in the mid-IR region from 5 to 25 μm confirms the quality of the grown CdZnTe single crystal. It is to be noted here that the theoretical transmission limit of $Cd_{0.9}Zn_{0.1}Te$ wafer in the mid-IR region is 63 %, however, the instrumental errors or the scattering losses from the surface may reduce the transmittance further [36].

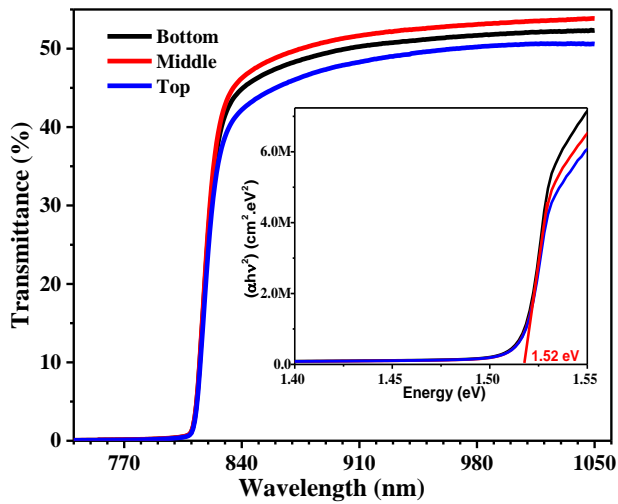

*Fig.7. NIR and mid-IR transmission spectrum of CdZnTe wafer*

### 3.3. I-V characteristics

I-V characteristics of a surface processed CdZnTe wafer is given in Fig. 8. Prior to I-V measurements, the CdZnTe wafers were gold coated in planar geometry by electroless method using AuCl3 solution. The I-V characteristics were measured using Keithley electrometer (6517B). As shown in the Fig. 8, the I-V curve is linear in the voltage range from -150 to +150 V. The resistivity of the wafer is calculated using the formula, $\rho = R*A/L$. Here, R is the resistance, A is the area of the electrode and L is the thickness of the sample. The resistance R is estimated using Ohm's law. The estimated resistivity of the CdZnTe wafer was determined to be ~ 6.6 x $10^{10}$ Ω.cm. This high bulk resistivity indicates good structural quality with minimal Cd vacancies. The CdZnTe wafers with high resistivity (>$10^{10}$ Ω-cm) were chosen for testing response to radiation dose.

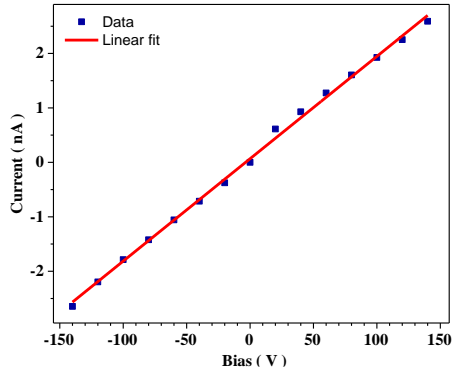

Fig. 8. I-V characteristics of CdZnTe wafer measured with planar gold electrodes.

### 3.4. Gamma radiation detection

The fabrication of the CdZnTe detector element involves multiple steps starting from dicing of ingot, mechanical or chemo-mechanical polishing, chemical etching, surface passivation, and electrode preparation. Each step in this process is crucial, since it can introduce significant surface defects which are detrimental to device performance [37]. Wet chemical etching by 0.1 to 2 % brominated methanol (Br-MeOH) is one of the common methods for minimizing surface damages created during mechanical polishing. Then, a mixture of $NH_4F$ and $H_2O_2$ was used for surface passivation. A quasi-hemispherical detector was prepared by gold electrode on the five sides of the CdZnTe crystal by the electroless process. The five sides of the crystal surfaces constitute the cathode. A small circular contact was deposited on the CdZnTe element top side, forming the anode. The sample was cased in an aluminium enclosure with a BNC connector and interfaced to a commercial charge-sensitive preamplifier.

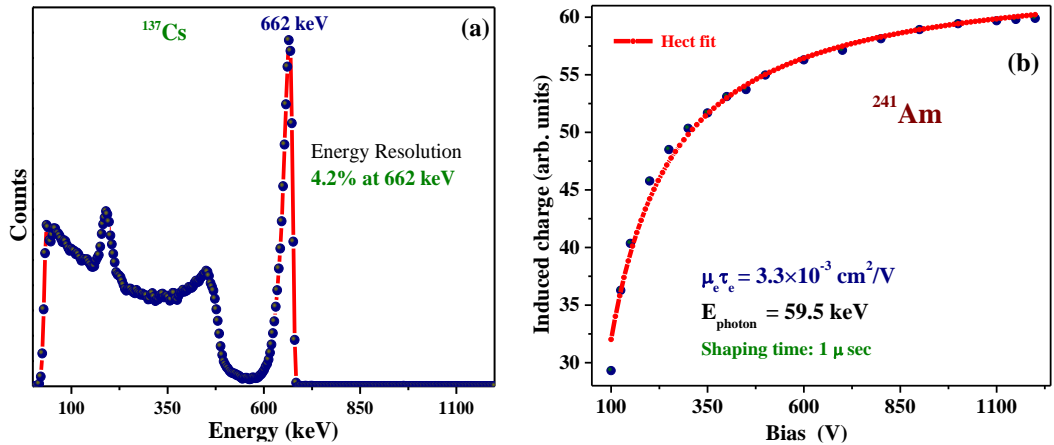

Fig. 9 (a) The $^{137}Cs$ gamma radiation spectrum recorded by in-house developed CdZnTe single crystal and (b) The induced charge as a function of applied bias on CdZnTe detector element

The gamma spectrum was measured using standard nuclear instrument modules consisting of spectroscopic amplifier and high voltage module. Spectrum acquisition was carried out using USB based multi-channel analyzer for 600 sec at a bias of 1200 V with a shaping time of 1 µs. The $^{137}$Cs gamma radiation source was used for the evaluation of spectroscopic performance. The recorded gamma spectrum is shown in Fig. 9a. The CdZnTe detector can resolve the photo peak of $^{137}$Cs at 662 keV with a resolution of 4.2 %. Here, the energy resolution of the in-house grown CdZnTe detector is little inferior as compared to the commertial CdTe and CdZnTe detectors. However, the energy resolution of the detector largely depends on the various surface preparation conditions. For example, the laser processing and ion plasma treatment of CdTe / CdZnTe wafers lead to the energy resolution down to ~ 0.5 % at 662 keV [38–40]. Besides, the photo peak measurements were also carried out at different biases ranging from 100 to 1200 V. The induced charge for the photo peak at 59.5 keV is plotted with respect to different biases as shown in Fig. 9b. The measured data is fitted with Hecht equation, and the deduced µτ product from the fit is ~ 3.3 x 10$^{-3}$ cm$^2$/V. This value is similar to the µτ values reported in the literature for CdZnTe crystals [41].

## 4. Conclusions

A process to grow nearly single-grain CdZnTe crystal in a 20 mm diameter quartz ampoule is demonstrated successfully using in-house designed THM crystal growth system. The growth parameters such as solvent to solute (Te/CdZnTe) ratio, temperature gradient, and growth rate were optimized to achieve 20 mm diameter large grain single crystal. At the laboratory scale from a starting charge of 100 g, at least 4 numbers of useable detector elements of dimensions 10 x 10 x 5 mm$^3$ can be reproducibly obtained from each growth experiment. X-ray diffraction, NIR optical transmittance and I-V characteristics affirm the growth of high-quality CdZnTe single crystals. The in-house developed CdZnTe detector exhibits an energy resolution of 4.2 % at 662 keV of $^{137}$Cs source with a µτ product of ~ 3.3 x 10$^{-3}$ cm$^2$/V. These gamma ray detector characteristics are a testimony to the successful growth of detector grade CdZnTe single crystals. Since the THM system is designed and developed in-house, it is easy to scale up for the growth of large size spectroscopic grade CdZnTe crystals in the future. Also, the in-house developed THM system can be extended to study other isomorphic compounds in the series such as CdZnTeSe, CdZnSe and CdMnTe.

**Acknowledgements**

The authors acknowledge the support and encouragement from Dr. B.Venkatraman, Director, IGCAR, Dr. R. Divakar, Director, MSG/IGCAR and Dr. R. Nagendran, Division Head, MSG/IGCAR. One of the authors, P.Vijayakumar, thanks IGCAR for the awards of Research Associate and Visiting Scientist fellowships.